\documentclass{vgtc}                          

\ifpdf
  \pdfoutput=1\relax                   
  \usepackage{graphicx}                
  \DeclareGraphicsExtensions{.pdf,.png,.jpg,.jpeg} 
\else
  \usepackage{graphicx}                
  \DeclareGraphicsExtensions{.eps}     
\fi%

\graphicspath{{figures/}{pictures/}{images/}{./}} 

\usepackage{microtype}                 
\PassOptionsToPackage{warn}{textcomp}  
\usepackage{textcomp}                  
\usepackage{mathptmx}                  
\usepackage{times}                     
\usepackage{cite}                      
\usepackage{tabu}                      
\usepackage{booktabs}                  

\def\_{\rule{.3em}{.15ex}}      

\newtheorem{definition}{Definition}
\newtheorem{property}[definition]{Property}
\newtheorem{proposition}[definition]{Proposition}
\newtheorem{lemma}[definition]{Lemma}
\newtheorem{theorem}[definition]{Theorem}
\newtheorem{corollary}[definition]{Corollary}

\newcommand {\mymarginpar}[1]{\marginpar{#1}}
\renewcommand {\marginpar}[1]{}

\newcommand {\rfig}[1]{Figure \ref{fig:#1}}

\newcommand {\bsec}[2]{\section{#1}
                       \label{sec:#2} }

\newcommand {\bsubsec}[2]{\mymarginpar{sec:#2}
                       \subsection{#1}
                       \label{sec:#2} }

\newcommand {\beq}[1]{
                      \begin{equation}
                      \label{eq:#1} }
\newcommand {\eeq}{\end{equation}}
\newcommand {\beqno}[1]{\begin{eqnarray}
                      \nonumber}

\newcommand {\eeqno}{ && \end{eqnarray}}

\newcommand {\bear}[1]{
                       \begin{eqnarray}
                       \label{eq:#1} }

\newcommand {\bearno}[1]{
                       \begin{eqnarray}
                       \nonumber}

\newcommand {\eear}{\end{eqnarray}}
\newcommand {\eearno}{\end{eqnarray}}

\newcommand {\btab}[1]{
                       \begin{table}
                       \centering
                       \begin{tabular}{#1}}
\newcommand {\etab}[3] {
                       \end{tabular}
                       \caption[#3]{#2}
                       \label{tab:#1}
                       \end{table}
                       \vspace{.1in}}

\newcommand {\btabular}[1]{\begin{center}
                       \begin{tabular}{#1}}
\newcommand {\etabular}{\end{tabular}
                       \end{center}}

\newcommand {\bdefin}[1]{\begin{definition}\label{def:#1}}
\newcommand {\edefin}       {\end{definition}}

\newcommand {\bpro}[1]{\begin{property}
                      \label{pro:#1} }
\newcommand {\epro}   {\end{property}}

\newcommand {\bprop}[1]{\begin{proposition}
                      \label{prop:#1} }
\newcommand {\eprop}       {\end{proposition}}

\newcommand {\blem}[1]{\begin{lemma}
                      \label{lem:#1}}
\newcommand {\elem}   {\end{lemma}}

\newcommand {\bthe}[1]{\begin{theorem}
                      \label{the:#1} }
\newcommand {\ethe}   {\end{theorem}}

\newcommand {\bcor}[1]{\begin{corollary}
                      \label{cor:#1} }
\newcommand {\ecor}   {\end{corollary}}

\newcommand{\hide}[1]{}

\onlineid{0}

\vgtccategory{Research}

\vgtcinsertpkg

\title{A Design Process of Visual Analytics Application using Conceptual Graph}

\author{Lei Shi\\ %
        \scriptsize Beihang University
}

\abstract{State-of-the-art visual analytics techniques in application domains are often designed by VA professionals over qualitative requirement collected from end users. These VA techniques may not leverage users' domain knowledge about how to achieve their analytical goals. In this position paper, we propose a user-driven design process of VA applications centered around a new concept called analytical representation (AR). AR features a formal abstraction of user requirement and their desired analytical trails for certain VA application, and is independent of the actual visualization design. A conceptual graph schema is introduced to define the AR abstraction, which can be created manually or constructed by semi-automated tools. Designing VA applications with AR provides a shared opportunity for both optimal analysis blueprint from the perspective of end users and optimal visualization/algorithm from the perspective of VA designers. We demonstrate the usage of the design process in two case studies.
} 

\begin{document}


\firstsection{Introduction}

\maketitle


\bsubsec{Background}{Background}

Visual analytics (VA) is coined as ``the science of analytical reasoning facilitated by interactive visual interfaces'' \cite{thomas2005visual}, or more specifically, ``combines automated analysis techniques with interactive visualizations'' \cite{keim2008visual}. In the new century, VA emerges as a promising methodology in the field of visualization and data science. Numerous successful VA applications have been presented in the annual VAST conference \cite{VASTConf}, contributing to key domains such as finance \cite{chang2007wirevis}, urban computing \cite{andrienko2008spatio}, machine learning \cite{wongsuphasawat2017visualizing}, and many others. Yet, state-of-the-art VA applications generally have indirect domain user involvement in their design process. Among six example papers on VA application \& system category provided in the VIS'20 website \cite{VASTPaperType}, five papers reported certain kind of user requirement or task analysis, the other paper mentioned little from the end user's perspective. In most of these works, raw input from end users is first summarized by VA experts (normally the authors) into list-based requirement/task, and then referenced in the VA design process.  Domain users do not directly contribute to the VA framework constructed for their applications. Notably, only one example paper planned the detailed analytical reasoning process of their VA solution in the design phase, for which the authors themselves were domain experts on the application of scientific literature analysis \cite{beck2015visual}.

Promoting domain user involvement in the VA design process to a level beyond qualitative requirement analysis adopted in classical visualization research can be beneficial. First, compared with visualization problems mostly dealing with fixed data types, VA applications are often built over more complex customer data where not all data input in their raw form are applied in solving the customer problem. Picking the right data source, selecting the most appropriate analysis direction all demand thorough understanding of user requirement and deep knowledge about the domain. This calls for closer involvement of end users in the VA design process. Second, existing requirement/task analyses in VA applications result in unstructured, textual lists as design guidelines. These guidelines normally vary a lot across VA applications. The final VA pipeline constructed in each application, though carefully follows these specific guidelines, is often seen as an one-time case-by-case design. It is difficult to reuse successful VA applications to automate the design in future. Third, from industry's perspective, the deficiency of domain user involvement in VA applications potentially slows down the adoption of the technology in the mass. By contrast, the growing popularity of software such as Tableau \cite{Tableau} and Microsoft Power BI \cite{PowerBI} witnesses the democratization of information visualization (InfoVis) techniques. Commodity users can freely drag-and-drop data items to gain much control of the visualization design, which may be one important reason of its success. Meanwhile, the widest spontaneous usage of VA technology outside the visualization community is seen in the deep learning field where the black-box models are analyzed and visualized for interpretability. Interestingly, many of these VA applications are constructed by machine learning experts who can gain full control of the design process as domain users. A similar call to embrace customers in the visualization community has been made many years before by Lorensen \cite{lorensen2004death}.

\bsubsec{Related Work}{Related}

\begin{figure*}[t]
\centering
\includegraphics[width=0.88\linewidth]{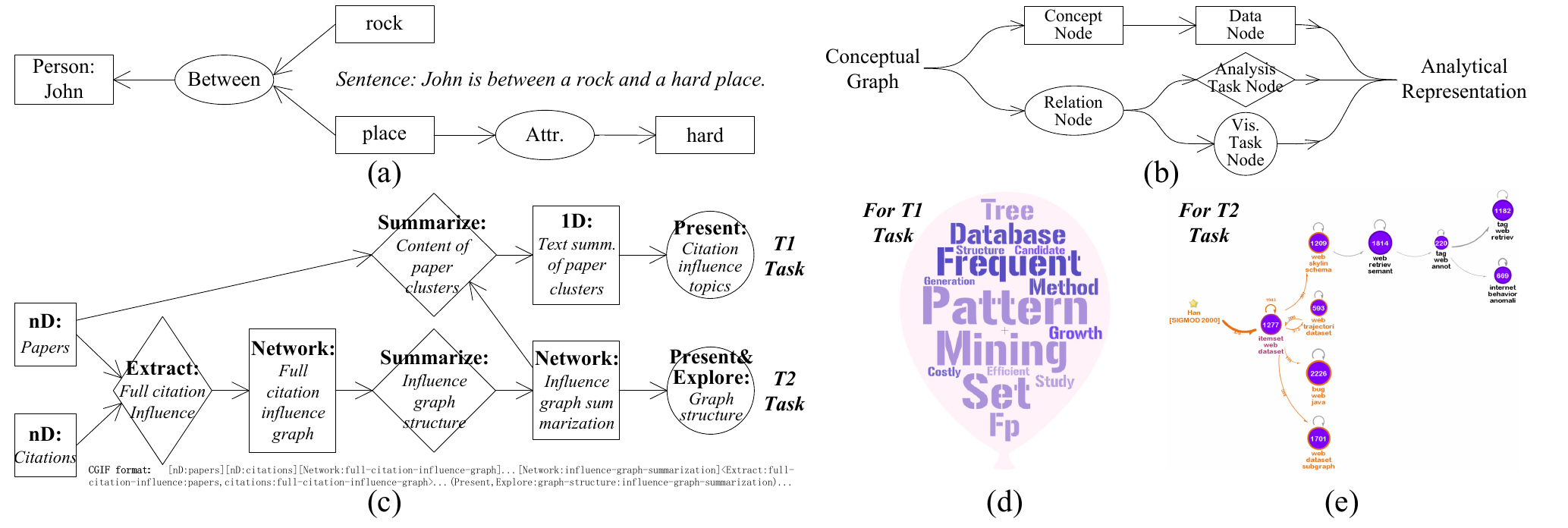}
\vspace{-0.2 in}
\caption{(a) Conceptual graph, (b) its adaptation for AR, and (c) an example of AR schema and (d)(e) visualization instances.}
\vspace{-0.2 in}
\label{fig:ConceptualGraph}
\end{figure*}

A central topic of visualization and VA research has been its design and implementation process. Card et al. proposed the InfoVis reference model \cite{card1999}, a standard pipeline to create interactive visualization from source data. Keim et al. illustrated a VA paradigm to integrate visualization with automatic data analysis methods for interactive decision support \cite{keim2008visual}. The visual information seeking mantra by Shneiderman suggested a golden rule in designing the exploration process with visualization \cite{shneiderman1996eyes}. Mckenna et al. presented a design activity framework composed of four critical activities for designing visualizations: understand, ideate, make, and deploy \cite{mckenna2014design}. All these high-level theories for visualization design might have been followed by many designers, but most of them do not consider the problem of getting direct involvement of end users in the VA design process. The theories help to modularize the implementation of each visualization stage, but provide little guidance in reusing VA solutions for customer problems. On implementing visualization software, a number of design patterns have been summarized in the literature \cite{heer2006software}\cite{chen2004toward}. Nevertheless, our work focuses on the VA design process, which is generally independent of its implementation.

More relevant to our study, the dataflow network used in scientific visualization (SciVis) \cite{dyer1990dataflow} provides low-level abstraction of data analysis and visualization pipeline for each application. By integrating with software toolkits \cite{schroeder2000visualizing} and user interfaces such as VisTrails \cite{bavoil2005vistrails}, the dataflow model allows greater user involvement in the visualization design. End users could fulfill their requirement and evaluate the visualization performance in the same interactive interface. Typically, SciVis experts behave as both end users and visualization designers in these applications. Compared with the dataflow model, we consider a different scenario in VA applications where domain users are normally not visualization and algorithm experts. These nonprofessionals may integrate their demand into the analytical pipeline of VA application through the dataflow-like model, but probably can not assemble the optimal visualization and algorithm choice, because of their lack of expertise in the VA field. In the InfoVis community, Chi and Riedl also proposed a data state model for visualization applications \cite{chi1998operator}, which can be seen as a dual form of the dataflow model. The data state model is mainly used to characterize visualization techniques with a focus on their interaction aspect. In contrast, we study the abstraction of user requirement/task in VA applications, which can be independent of their detailed visualization and interaction design.

Related to our work, the user tasks accomplished in visualization were intensively discussed in the literature. Shneiderman proposed a task by data type taxonomy which includes seven data types (1D, 2D, etc.) and seven visualization tasks (overview, zoom, etc.) \cite{shneiderman1996eyes}. Stasko summarized ten low-level analysis tasks in understanding data with visualization \cite{amar2005low}, which were compiled over existing low-level task taxonomies \cite{wehrend1990problem}. Brehmer and Munzner studied both low-level and high-level visualization tasks and proposed a multi-level typology to orchestrate the task design space \cite{brehmer2013multi}. Similar researches have also been conducted on the taxonomy and design space of visualization tasks with the multidimensional approach \cite{schulz2013design}, in cubic views \cite{rind2016task}, from the perspective of user interaction \cite{heer2012interactive}, and for certain data types \cite{NetworkTask}\cite{valiati2006taxonomy}. These existing studies on classifying visualization tasks are orthogonal to our goal of promoting user involvement to fulfill their tasks in VA applications. In fact, we borrowed some of the high-level task taxonomy in defining the task node in the representation of user requirement for visual analytics. Notably, our work is also motivated by Munzner's nested model for visualization design \cite{munzner2009nested}, which decouples domain problem characterization (user requirement), data/operation abstraction (analysis and visualization task), and visualization design (user performance). The VA design process introduced here partially fulfills Munzner's nested model in the first and second levels of abstraction.

\bsubsec{Two-Stage VA Design Process}{Proposal}

Getting user involvement in VA design brings the immediate challenge of having unprofessional visualizations. The main idea of this work is to allow structured input of domain requirement/knowledge from end users, while imposing the separation of satisfying user requirement from optimizing visualization performance in VA applications. To achieve this idea, a two-stage VA design process is proposed. In the first stage, an abstraction of end user's visual analysis trail is described in the form of data analysis and visualization tasks. The abstraction is independent of the visualization design so that end users can get involved in this stage to the largest extent without caring about the genre and performance of final visualization design. In the second stage, the first-stage abstraction is translated by VA designers into the classical VA pipeline, through optimally selecting data analysis algorithms and visualizations, e.g. by the design activity framework of Mckenna et al. \cite{mckenna2014design}. The optimal VA design can not only be used in the current application, but also recorded in VA knowledge base indexed by structured requirements, which facilitates reuse/automation of VA applications.

Provisioning the two-stage VA design process is more difficult than envisioning it. How to isolate user requirement from their performance when the two are often entangled in visual analytics? How to design the abstraction in the first stage to minimize the gap for end user's direct involvement while ensuring formality for faithful translation into optimal VA designs in the second stage? We make two contributions to address these challenges.

\begin{itemize}
 \item The new concept of analytical representation (AR) is introduced as the abstraction of user requirement and analytical trails in VA applications. We propose a formal method using conceptual graph to describe the intension of AR, and using complementary visualization and algorithm instances to illustrate its extensions. End users could precisely integrate their desired visual analytics tasks into AR and offer VA designers a detailed blueprint for constructing VA applications.
 \item The VA design process with AR is presented and evaluated. Case study results show that by the new design process, visual analytics in our sample applications are made both adaptive and dedicated to solving end user's domain problems. In the cases, the design process generates VA pipelines that are compatible to existing reference models.
 \end{itemize} 
\bsec{Analytical Representation}{AR}

\bsubsec{Definition}{Definition}






Analytical representation is the visualization-independent abstraction of relevant data and processing for certain visual analytics application. Primarily, AR is defined for and used in VA scenarios where domain problems are solved by combining automated data analysis and human-in-the-loop visualization. The main body of AR is the abstraction of required trails of data analysis and visualization from the view of end users. Most importantly, the abstraction is independent of the actual visualization design and data analysis algorithms. In a sense, AR could be the mental map of end users on how to solve a domain problem using the combination of data analysis and visualization, though he may not know and need not know the detailed method in each step of visual analysis. \rfig{ConceptualGraph}(c) gives an example of AR schema for citation influence analysis by linked high-level task abstractions on data analysis and visualization.

Empowered by the concept and tool to create AR, end users could focus on the requirement and domain knowledge for VA applications, without considering the performance using particular visualization design. The AR abstraction can be exemplified by automatically generated visualization instances, which help end users early evaluate the effectiveness of VA applications at the design stage. End users could iterate AR blueprint using evaluation result as feedback. AR itself also serves as a quantitative index of knowledge base for VA best practices. Building such knowledge base could push forward the reuse, automation, and finally wider adoption of VA technology.

\bsubsec{AR Schema by Conceptual Graph}{CG}


To achieve formalism in analytical representation, we propose to use conceptual graph (CG) to define the schema of AR. Historically, CG is first invented to represent database schema and natural language (see an example in \rfig{ConceptualGraph}(a) ) \cite{sowa1983conceptual}, and later extended to support a wider scope of knowledge representation \cite{chein2009graph}. AR, defined as the representation of end user's requirement and domain knowledge for certain VA application, is also a kind of knowledge representation. This is the main reason we apply CG as the schema of AR. Additionally, the expressive power of labeled graph notation in CG allows the level of abstraction in AR definition.


There are necessary adaptations from the initial definition of CG to the schema of AR, which is illustrated in \rfig{ConceptualGraph}(b). CG is originally composed of the concept nodes and the relation nodes. The concept node is altered to the data node in AR (boxes in \rfig{ConceptualGraph}(c)). Each data node has a data type and a data instance. The relation node is altered to two classes of nodes: the analysis task node (diamonds in \rfig{ConceptualGraph}(c)) and the visualization task node (circles in \rfig{ConceptualGraph}(c)). Each task node has a task type and a task target. Directed arcs connecting nodes in AR represent data flows.

The key concept of AR is the typed node representation. Each class of nodes can define its own type ontology, i.e., a multi-level category (lattice) of types. For data nodes, we use the taxonomy by Shneiderman as the top-level category for data types \cite{shneiderman1996eyes}: 1D, 2D, 3D, nD, temporal, tree, network. A different taxonomy can also fit in depending on the usage scenario of AR. For visualization task nodes, we advocate the use of high-level task taxonomy as its type category, as low-level task nodes can make the abstraction of AR overly detailed and less reusable. In this work, we apply the high-level task typology by Brehmer and Munzner \cite{brehmer2013multi}: present, produce, explore, identify, etc. Again, using a different high-level visualization task taxonomy is possible. Analysis task nodes are not our focus, any ontology of data analysis algorithms will work.

In the example of \rfig{ConceptualGraph}(c), a visual analysis trail on the citation influence of a scientific paper is represented by the adapted CG form of AR schema. As end users expect, a full citation influence graph of the paper should be extracted and summarized. Then s/he wants to visualize and interactively explore the influence graph summarization (T2 visualization task). Also, the user wants to be presented a text summarization of paper clusters in the graph to understand the citation influence content (T1 visualization task). Inherited from CG notations, the schema of AR can also be represented in a textual format known as Conceptual Graph Interchange Format (CGIF). The bottom line of \rfig{ConceptualGraph}(c) gives part of this format for the example. The CGIF of AR can be analyzed by symbolic computation, similar to knowledge reasoning and inference on CGs.



CG is an abstract schema of AR. Beyond the abstraction, end users also need a preview and early evaluation of the VA design over the AR schema. We propose to link each visualization task node in AR with at least one instance. The instances can be retrieved from the knowledge base of VA designs indexed by AR graphs and nodes. With sufficient VA applications input to the knowledge base, the instance can be automatically recommended as the design of the best-matched or most-representative task node in the knowledge base for the current node. \rfig{ConceptualGraph}(d)(e) presents artificial instances for the two visualization task nodes in \rfig{ConceptualGraph}(c). A standard tag cloud visualization is mapped to T1 visualization task node \cite{viegas2009participatory}. A specialized visualization for clustered citation graph \cite{shi2015vegas} is selected for T2 task node. If necessary, algorithm instances of the analysis task node can also be delivered to end users for evaluation.

\bsubsec{Comparison to Other Concepts}{Comparison}

AR is related to many existing concepts, but there are clear distinctions. Data representation is the class of methods primarily used to prepare data for machine computing and is not designed for visualization purpose. Visual representation, on the other hand, describes the mapping from data to visualization, but does not consider domain-specific VA tasks. AR can be seen as a concept between data and visual representation with the focus to abstract high-level data analysis and visualization tasks for certain VA application.

In the knowledge representation community, ontology and knowledge graph are widely adopted. They are used for the representation of subjects and entities in one field, while AR is used as the representation of VA tasks in certain application. Graphical models such as neural networks are designed to represent data distributions, which are special types of data representation. They are not at the same level as AR in the pipeline of visual data analysis.

\bsec{Designing Visual Analytics with AR}{Process}

\begin{figure}[t]
\centering
\includegraphics[width=0.99\linewidth]{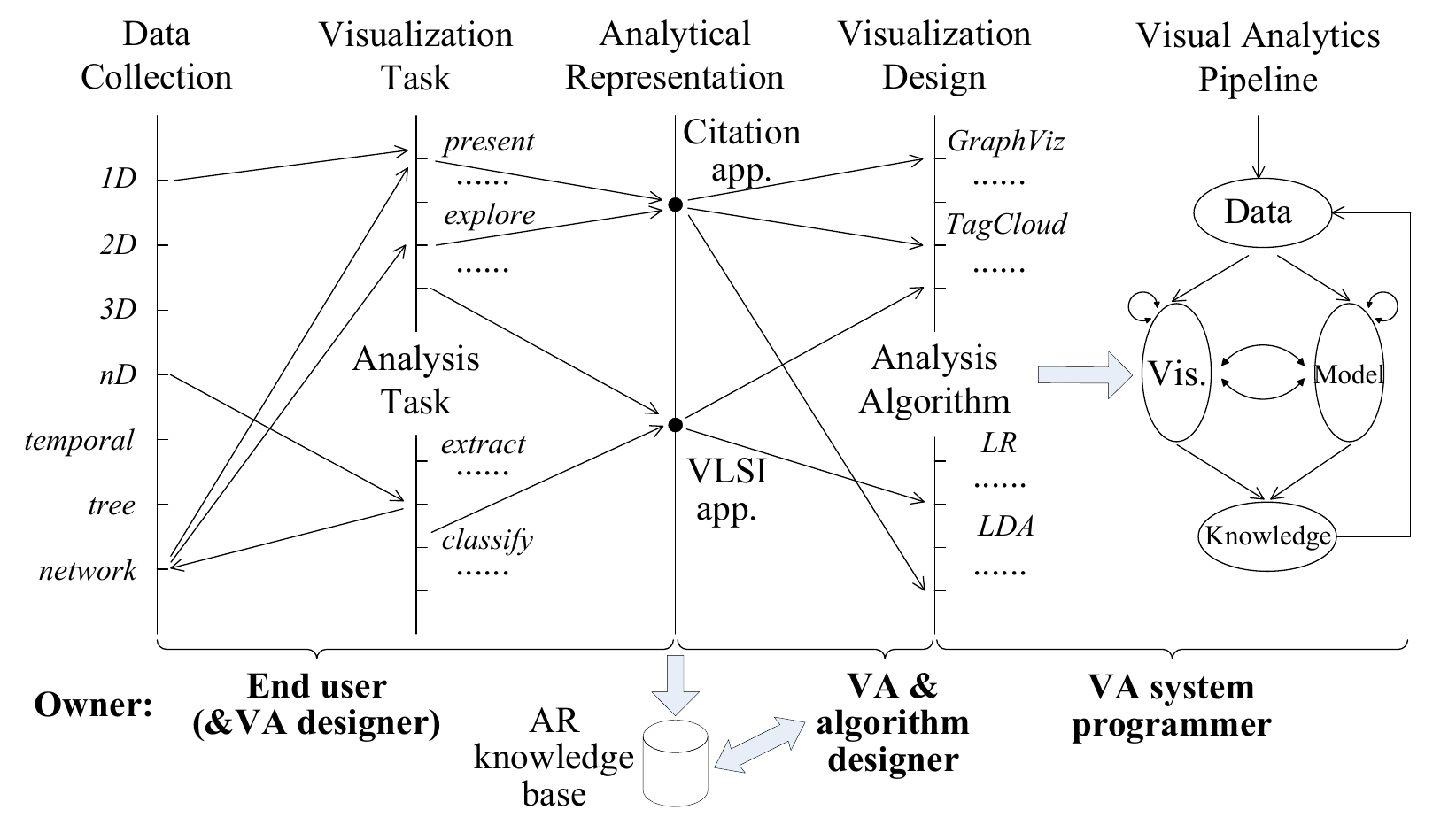}
\vspace{-0.2 in}
\caption{The design process of VA applications with AR.}
\vspace{-0.15 in}
\label{fig:DesignProcess}
\end{figure}

\rfig{DesignProcess} suggests a reference model for designing VA applications with AR. In the first half of the model before AR, user requirement is analyzed to characterize VA tasks separately on data analysis and visualization. These tasks and available data input for the application are connected according to end user's desired visual analytics trail of the application. We note that end users can compose an AR of the application solely by themselves, but the participation of VA designers in this stage is highly recommended. Users may not be familiar with the category system of VA tasks and the capability of related visualizations and algorithms. An alternative AR design approach is to automatically extract AR blueprint from visualization sketches of the end user. The automated approach can be highly challenging and is not the focus of this work.

In the second half of the model in \rfig{DesignProcess}, VA designers and algorithm designers create the actual visualizations used in the VA application and select appropriate algorithms for the data analysis. These visualizations and algorithms should cover all task nodes specified in the AR blueprint. The AR-based VA design process exhibits a clear advantage in that designers may pick visualizations and algorithms according to their experience and expertise, and can also leverage the AR knowledge base storing the design of qualified VA applications. The knowledge base can suggest or even automate the design of certain AR task nodes using existing best practices. Another advantage of the proposed VA design process lies in that the output is still a standard VA pipeline. Every stage is already decided in the design process and can be implemented by VA system programmers in the same way with previous VA applications. 
\bsec{Case Study}{Eva}

\begin{figure*}[t]
\centering
\includegraphics[width=\linewidth]{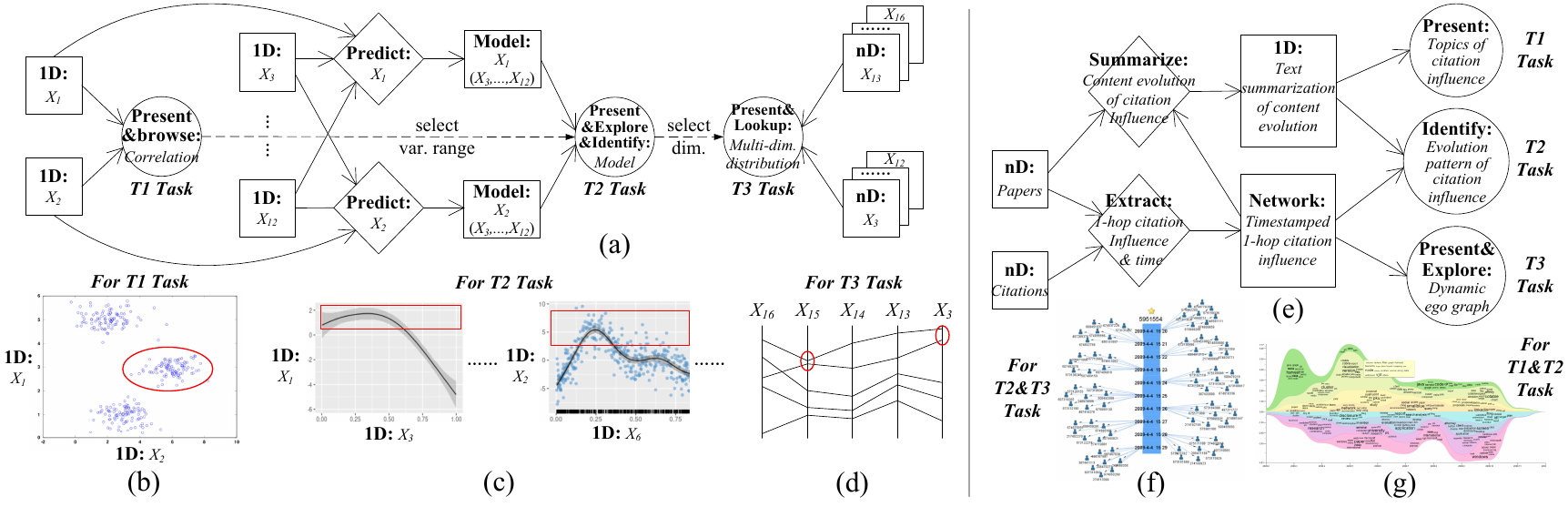}
\vspace{-0.3 in}
\caption{VLSI chip production case: (a) AR schema; (b)(c)(d) visualization instances. Citation analysis case: (e) AR; (f)(g) visualization instances.}
\vspace{-0.13 in}
\label{fig:VLSI}
\end{figure*}

We present two case studies to demonstrate the usage of AR-based design process for VA applications. In the first case study, we show how the integration of domain knowledge and user requirement in AR can allow clearer path towards user's analytical goal. The background is the well-known application of parallel coordinates visualization (PCP) by Inselberg in analyzing the production dataset of VLSI chips \cite{inselberg1997multidimensional}. The dataset is composed of 16 variables: $X_1$ and $X_2$ are yield and quality of each production batch; $X_3\sim X_{12}$ are the number of defects in ten types; $X_{13} \sim X_{16}$ are physical parameters of each batch. Through iterative navigation of the dataset with PCP, several data patterns are found to help maintain both high batch yield and quality, including a small range of $X_3$, $X_6$, and $X_{15}$.

While we applaud for the detective ingenuity of the author, there is also doubt about whether ordinary users can discover the same pattern with PCP. In an alterative approach using visual analytics and AR, users start from their requirement, i.e., to obtain high yield and quality in a batch. By VLSI literature \cite{koren1990fault}, it is also common sense for domain users that yield and quality ($X_1$,$X_2$) can be modeled by \#defects ($X_3 \sim X_{12}$), which are further determined by physical parameters ($X_{13} \sim X_{16}$). With this domain knowledge, we believe end users can achieve the AR design of \rfig{VLSI}(a), albeit certain variations among users. In their design, $X_1$ and $X_2$ are first visually correlated to clarify the analytical goal: what is the desired high range of yield and quality (see the red circle in the scatterplot instance of \rfig{VLSI}(b))? To answer which defects affect the most on high yield/quality, predictive models can be built and visually analyzed. In simplified instances by the regression model and visreg visualization \cite{breheny2017visualization} (\rfig{VLSI}(c)), users could easily identify $X_3$ and $X_6$ as key types of defects for chip yield and quality where $X_1$ ($X_2$) changes significantly with $X_3$ ($X_6$) outside the selected high range in red. Adding these dimensions to lookup the desired physical parameters  ($X_{13} \sim X_{16}$), the PCP instance in \rfig{VLSI}(d) suggests a small range of $X_{15}$, the same pattern detected by Inselberg.

The second case study shows that the VA design process with AR helps to distinguish between different user requirement for the same domain problem. In our previous work, we have developed a VA system for citation influence analysis, whose AR schema has been used as the example in \rfig{ConceptualGraph}(c). During the research cycle, we also received from reviewers and other academic users about different opinions of our work. Notably, they consider the citation influence of a paper as the direct 1-hop citation network from the original paper, instead of the full citation network measuring the influence on a field. In addition, they also want to understand the temporal dynamics of the influence, which is not considered by us. By the AR-based design, the new user requirement can be satisfied by simply revising the AR blueprint. As shown in \rfig{VLSI}(e), the extract task node is changed to analyze 1-hop citation influence. The summarize influence graph node is removed as there may not be many nodes in the 1-hop citation influence graph. The time dimension is incorporated in most task nodes and a new visualization task node is added to identify evolution patterns from citation influence (T2). The visualization instances in \rfig{VLSI}(f)(g) clearly reflect the different user requirement in comparison to the previous instances (\rfig{ConceptualGraph}(d)(e)). Using AR-based VA design process, both similar and divergent user requirement can be represented and executed, which effectively enhances the reusability of VA applications. 
\bsec{Discussion}{Discussion}


As a preliminary framework, the current AR design process does have limitations. First, AR specifies analytical trails in VA applications, but it is not designed for explorative visual analytics in which end users have little knowledge about the domain and/or do not have clear analytical goal. The deficiency can be alleviated by adding a feedback mechanism to the AR design process. An initial AR can be constructed from general-purpose analysis tasks. After gaining new knowledge and potential analytical goals in the domain, users can update the AR design by translating the new knowledge/goal into visual analytics trails. Repeating the design process helps to solve domain problems with VA techniques.


Second, there are threats in constructing AR. End users may not have the capability to summarize their requirement into appropriate task nodes and an accurate AR schema. While VA designers can certainly provide support through direct communication, we believe that the end users designing VA applications with AR should have moderate skills in data analytics, though they may not be visualization experts. Unlike automatic data analysis applications such as data mining, visual analytics inherently require end users to understand and manipulate the data analysis process. For these users, composing AR blueprints should be possible with training. In addition, semi-automated AR construction tools providing taxonomies and instances of visualization/analysis tasks will further reduce the barrier for the proposed VA design process.


Third, as a position paper, many remaining work need to be carried out along the direction. The existing taxonomies of data types, analysis and visualization tasks should be studied for the best fit in AR design. The design automation tool for AR will also expedite the usage of AR-based VA applications. The theory and implementation of AR knowledge base is another promising future work towards democratization and reusability of VA applications. 
\bsec{Conclusion}{Conc}
Synthesizing end user's requirement and their domain knowledge is challenging in existing VA applications. In this position paper, we show that it is possible to formally abstract this critical information by a new concept called analytical representation. AR adopts the conceptual graph as its schema and can be evaluated through extended visualization instances. Designing VA applications with AR enjoys three advantages. First, AR could be used as a blueprint for solving specific domain problem following the analytical trails recommended by end users. Second, the visualization-independent nature of AR allows the optimization of visualization and algorithm designs in the application through separate effort of VA/algorithm designers. Third, the abstraction enables quantification of VA designs and the establishment of VA knowledge base, so that design automation and reuse of VA applications can be expected. We also present two case studies to demonstrate the desired advantages of the new design process.

\bibliographystyle{abbrv-doi}

\bibliography{AR_VIS}
\end{document}